\documentclass[a4paper,10pt,twoside]{cpc-hepnp}

\usepackage{multicol}
\usepackage{graphicx}
\usepackage{booktabs}
\usepackage{amssymb,bm,mathrsfs,bbm,amscd}
\usepackage[tbtags]{amsmath}
\usepackage{lastpage}

\begin{document}


\footnotetext[0]{Received 11 April 2016}

\title{A Common Optical Potential for $^{4}$He+$^{12}$C at Intermediate Energies  \thanks{Supported by National Natural Science Foundation of China (Grant No.11205036) and the Fundamental Research Funds for the Central Universities of China (Grant No.HEUCF101501) }}
\author{%
      Li-Yuan Hu$^{1}$%
\quad Yu-Shou Song$^{1;1)}$\email{songyushou80@163.com}%
\quad Ying-Wei Hou$^{1}$
\quad Hui-Lan Liu$^{1}$
}
\maketitle
\address{%
$^1$ Fundamental Science on Nuclear Safety and Simulation Technology Laboratory¡AHarbin Engineering University, Harbin 150001\\
}
\begin{abstract}
A common optical potential for $^4$He+$^{12}$C at intermediate bombarding energies, which is essential in analyzing exotic nuclei with $^4$He clusters, was obtained based on the S\~{a}o Paulo potential (SPP). Among systematic optical potentials for $^4$He+$^{12}$C, this potential has the merit of using a fixed imaginary part of Woods-Saxon form. By optical-model calculations, this potential reproduced the experimental elastic scattering angular distributions of $^4$He+$^{12}$C well within the energy range of 26\,$A$--60\,$A$ MeV. It was also applied successfully in calculations of the breakup reactions of $^6$Li+$^{12}$C and $^6$He+$^{12}$C with a three-body continuum discretized coupled-channels method. 

\end{abstract}


\footnotetext[0]{\hspace*{-3mm}\raisebox{0.3ex}{$\scriptstyle\copyright$}2016
Chinese Physical Society and IOP Publishing Ltd}%

\begin{multicols}{2}
As one of the simplest nuclei, $^4$He is also an important cluster in exotic nuclei, such as $^{6,8}$He~\cite{b1,b2} and $^{12,14}$Be~\cite{b3,b4}. The breakup reaction of these nuclei on $^{12}$C target is an effective way to extract their cluster information~\cite{b5}. The continuum discretized coupled-channels (CDCC) method~\cite{b6} is traditionally employed to handle this reaction calculation, where the optical potential (OP) between $^4$He and $^{12}$C target is necessary. Unfortunately, the experimental data of elastic scattering angular distribution (ESAD) for some energy used to extract OPs does not exist. Alternatively, we can choose the OP of an adjacent energy, which is in rough approximation.  
The OP between $^4$He and $^{12}$C is supposed to be studied systematically.  
A few global OPs~\cite{b7,b8} have been presented based on existent experimental data. However, satisfactory ESADs have not been given for a particular projectile-target system at some energies. 
There are also some microscopic OPs for $^4$He~\cite{b9,b10,b11} obtained by double- or single-folding approaches. The real and imaginary parts are both determined, while their strengths need to be modified for different energies. 
Guo {\it et al.} proposed a microscopic OP by employing the Green function method. A substantial difference was observed between the calculated ESADs and the experimental results at larger angles~\cite{b12}.

The S\~{a}o Paulo potential~\cite{b13}, a nonlocal double-folding model for the heavy-ion nuclear interaction, is a good choice for the systematic study of the OP. In Ref.~\cite{b14}, it was used to develop a parameter-free OP with the nuclear component in the form of $V_{\rm SPP}+i0.78V_{\rm SPP}$. This OP reproduces the ESADs of seven heavy-ion systems (for example, $^{12}$C+$^{12}$C and $^{16}$O+$^{208}$Pb) at different energies fairly well without any adjustments. However, it is not suitable for $^4$He+$^{12}$C due to the fact that the experimental data of refractive $^4$He-nucleus systems require a different shape for the absorptive potential~\cite{b15}.

Based on the SPP,  we analyze the $^4$He+$^{12}$C elastic scattering at intermediate energies with the optical model (OM). 
The SPP is selected as the real part of the nuclear component of the OP, and an adjustable Woods-Saxon (WS) form is assumed for the imaginary part. 
The existing experimental ESADs obtained from EXFOR~\cite{b16} are fitted in the OM calculations by varying the parameters of the imaginary potential. The best-fit parameters are determined and the ESADs calculated with these parameters are in quite good agreement with the experimental data.
Furthermore, a fixed imaginary part is found to be possible to construct a common OP (COP) for $^4$He+$^{12}$C at energies ranging from 26\,$A$\,MeV to 60\,$A$\,MeV. This COP is tested in the CDCC calculations for the reactions of $^6$Li+$^{12}$C and $^6$He+$^{12}$C. The ESADs of the former are reproduced well. The ESAD of the latter system is underestimated by the calculation at larger angles, which is consistent with Keeley {\it et al.}~\cite{b17}. The calculations based on OM and CDCC are performed by the code FRESCO~\cite{b18}.

As usual, the OP between $^4$He and $^{12}$C is composed of a Coulomb part and a complex nuclear part. The nuclear potential $U_{\rm N}$ is expressed as 
\begin{equation}
\label{eq:OP}
U_{\rm N}(R)=V(R)+iW(R). 
\end{equation}
The SPP adopting a zero-range approach~\cite{b13} is chosen as the real part ($V$).
\begin{equation}
\label{eq:SPP}
V_{\rm SPP}(R)=V_{0}\int\rho_{\rm m1}(\bm{r}_{1})\rho_{\rm m2}(\bm{r}_{2})\delta(\bm{R}-\bm{r}_{1}+\bm{r}_{2})\\
e^{-4v^2/c^2}\,d\bm{r}_{1}\,d\bm{r}_{2},
\end{equation}
where the effective interaction is in the form of a delta function and $V_{0}$ =$-$456 MeV$\cdot$fm$^3$. The matter density $\rho_{\rm m}$ is described by the two-parameter Fermi (2pF) distribution~\cite{b13}:
\begin{equation}
\label{eq:rho}
\rho_{\rm m}(r)=\frac{\rho_{\rm m0}}{1+\exp({\frac{r-R_{\rm m0}}{a_{\rm m}}})}.
\end{equation}
where $\rho_{\rm m}$ of $^{12}$C is determined by the SPP systematics for the densities of heavy ions ($R_{\rm m0}$=2.159\,fm, $a_{\rm m}$=0.56\,fm)~\cite{b13}. $^4$He is too light to be covered by this systematics, thus specific values of $R_{\rm m0}$=1.162\,fm and $a_{\rm m}$=0.42\,fm are used~\cite{b19}. The exponential term includes the Pauli nonlocality between the projectile and the target nuclei. 
The energy dependence of the potential is implied in the relation between the local relative velocity $v$ and the kinetic energy~\cite{b19}. 
The imaginary part ($W$) uses a WS form factor with the radius $R_{\rm I0}$=$r_{\rm I0}A_{\rm t}^{1/3}$. 
For the Coulomb interaction $U_{\rm C}$, we assume that the Coulomb radius parameter $R_{\rm C}$=1.3$A_{\rm t}^{1/3}$\,fm.

With this OP, experimental ESADs of $^4$He on $^{12}$C target at different energies~\cite{b20,b21,b22,b23,b24} (Table \ref{Tab:001}) are analyzed in the OM framework. The adjustable parameters in the imaginary part, i.e., $W_{0}$, $r_{\rm I0}$ and $a_{\rm I}$, are determined by fitting the experimental data. Good agreement between the theoretical calculations and the experimental data is achieved (Fig.\,\ref{fig:001}).
The best-fit $W_{0}$, $r_{\rm I0}$ and $a_{\rm I}$ (set 1) fall into the ranges of 14.2-16.6\,MeV, 1.70-1.80\,fm and 0.50-0.60\,fm, respectively, except the data at 30\,$A$\,MeV and 34.75\,$A$\,MeV. 
Certain elastic scattering data may lead to potentials with different parameters within the error limit. To keep consistency with the parameters of the other data sets, the diffuseness $a_{\rm I}$ of the imaginary potential derived from the data at 30\,$A$\,MeV is reduced to 0.55\,fm, the middle value of the diffuseness fitted by the other data sets. The fitting procedure is redone for this data. The obtained values of $r_{\rm I0}$ and $W_{0}$ enter the ranges of the other data (see Table \ref{Tab:001}) as well. This can be understood as the correlation among $r_{\rm I0}$, $a_{\rm I}$ and $W_{0}$ within certain limits~\cite{b25}. We carry out the same thing for the data at 34.75\,$A$\,MeV except that $W_{0}$ was tuned manually to make the fit  better at larger angles. The two new sets of parameters are tagged set 2 in Table \ref{Tab:001}.

As listed in Table \ref{Tab:001}, there is no explicit dependence of the imaginary-part parameters on energy, and their values of each data set are close to each other. This makes it possible to use a fixed-WS imaginary part for these data sets. To obtain its appropriate parameters, we perform a WS-function fit based on the points given by the discretization of these WS functions plotted in Fig.\,\ref{fig:002} (set 2 of data sets at 30\,$A$\,MeV and 34.75\,$A$\,MeV, and set 1 of the other data sets). The outcomes $W_0$=15.48\,MeV, $r_{\rm I0}$=1.760\,fm and $a_{\rm I}$=0.552\,fm. With this fixed imaginary part and the SPP used as the real part, the COP is constructed and applied to the OM calculations of $^4$He+$^{12}$C. The calculated ESADs agree with the experimental data well without adjustments (Fig.\,\ref{fig:001}) and the $\chi^2/N$ values only become slightly larger (Table \ref{Tab:001}).

\vskip -2mm
\begin{center}
\includegraphics[width=\linewidth]{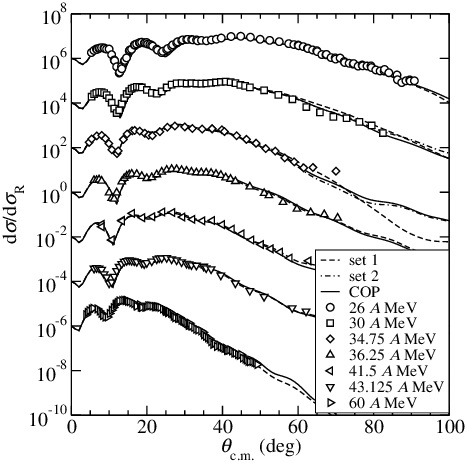}
\figcaption{\label{fig:001}Comparisons between OM calculations with WS imaginary part and experimental ESAD data of $^4$He+$^{12}$C at different energies~\cite{b20,b21,b22,b23,b24}. The data set at 36.25\,$A$\,MeV is on the real scale, and the others are offset by factors of 100 for good views. }
\end{center}
\vskip -5mm

\begin{table*}
\begin{center}
\tabcaption{\label{Tab:001}The best-fit parameters of the imaginary parts at different energies. Here $\chi_{\rm best}^2/N$ and $\chi_{\rm COP}^2/N$ correspond to the best-fit OPs and the COP, respectively. The last column gives the references of the experimental data.}
\footnotesize
\begin{tabular}{cccccccc}
\toprule
$E/A$ (MeV) & Set & $W_{0}$ (MeV) & $r_{{\rm I}0}$ (fm)& $a_{\rm I}$ (fm) & $\chi_{\rm best}^2/N$ & $\chi_{\rm COP}^2/N$ & Reference\\
\hline
26 & 1 & 16.63 & 1.707 & 0.550 & 31.161 & 32.652 & Ref.\cite{b20} \\
30 & 1 & 19.44 & 1.595 & 0.673 & 20.978 & $-$ & Ref.\cite{b21} \\
   & 2 & 14.77 & 1.799 & 0.550 & 23.959 & 24.421 &  \\
34.75 & 1 & 20.54 & 1.596 & 0.644 & 362.158 & $-$ & Ref.\cite{b22} \\
      & 2 & 16.00 & 1.748 & 0.550 & 494.424 & 521.180 &  \\
36.25 & 1 & 15.33 & 1.788 & 0.539 & 11.106 & 12.945 & Ref.\cite{b21} \\
41.5 & 1 & 14.23 & 1.795 & 0.504 & 13.149 & 18.881 & Ref.\cite{b23} \\
43.125 & 1 & 15.44 & 1.778 & 0.544 & 12.072 & 13.260 & Ref.\cite{b21} \\
60 & 1 & 15.99 & 1.708 & 0.608 & 2.760 & 6.127 & Ref.\cite{b24} \\
\bottomrule
\end{tabular}
\end{center}
\end{table*}

\vskip -25mm
\begin{center}
\includegraphics[angle=0, width=\linewidth]{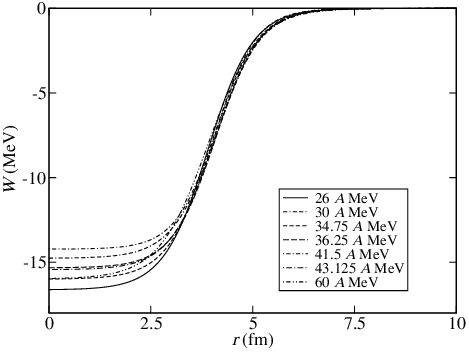}
\figcaption{\label{fig:002}The imaginary parts of WS form of OPs obtained by fitting experimental ESADs of $^4$He+$^{12}$C at different energies (set 2 of data sets at 30\,$A$\,MeV and 34.75\,$A$\,MeV and set 1 of the other data sets).}
\end{center}

The COP is utilized in the OM calculations of $^4$He+$^{12}$C at other energies as well to analyze the applicable energy range. 
Using the COP in the reaction at 27.5\,$A$\,MeV~\cite{b26}, the experimental ESAD is reproduced successfully as those data sets in Fig.\,\ref{fig:001}. At 12.625\,$A$\,MeV~\cite{b27}, 13.525\,$A$\,MeV~\cite{b28} and 16.25\,$A$\,MeV~\cite{b29}, the calculated ESADs are acceptable before 30 degrees (c.m. system), however, it underestimates the experimental data at larger angles. At 96.5\,$A$\,MeV~\cite{b30}, there is also an evident discrepancy between the computed ESAD and the experimental one.  
As a result we infer that this COP is appropriate within the energy range roughly from 26\,$A$\,MeV to 60\,$A$\,MeV.

Further, the COP is also used in the three-body CDCC calculations of $^6$Li+$^{12}$C at 28.1\,$A$\,MeV~\cite{b31}, 35\,$A$\,MeV~\cite{b32} and 53\,$A$\,MeV~\cite{b33}, and $^6$He+$^{12}$C at 38.3\,$A$\,MeV~\cite{b34}. The breakup couplings to the elastic scattering are considered. Two-body cluster structures (core+valence) are assumed for projectiles $^6$Li=$^4$He+$d$\,~\cite{b35} and $^6$He=$^4$He+2$n$\,~\cite{b1}. In these calculations, the OPs for core+target and valence+target are needed. The COP is used as the interaction between $^4$He and $^{12}$C. And the global OP proposed in Ref.~\cite{b36} was used for $d$+$^{12}$C and 2$n$+$^{12}$C. 

The spin of the valence $d$ of $^6$Li is ignored, which leads to the degeneration of three original $l$=2 resonances at about the relative energy $E_{\rm rel}$=2\,MeV~\cite{b35}. 
The binding potential (BP) of $^4$He+$d$ is of WS form with a radius $R_{\rm 0}$=1.9\,fm and a diffuseness $a_{\rm 0}$=0.65\,fm~\cite{b17,b37,b38}. For the ground state, a depth of 77.46\,MeV is used in the BP to give the binding energy of 1.471\,MeV~\cite{b37}. The depths for the non-resonant continuum and the resonance are 77.5\,MeV and 79.44\,MeV, respectively. 
In the construction of the model space, relative s-, p- and d-waves between $^4$He and $d$ are included. The $^4$He+$d$ continuum is discretized in the momentum space with a step $\Delta\,k$=0.15\,fm$^{-1}$ up to 1.5\,fm$^{-1}$. The resonance of $l$=2 is specially set as a single bin with a width of 3\,MeV. These bin states are generated using the mid-point method.
To reproduce the experimental ESADs better, renormalization factors $N_{\rm R}$ for the real parts of nuclear potentials are introduced~\cite{b38} (Table \ref{Tab:002}), and good results are obtained (Fig.\,\ref{fig:003}). 

\begin{center}
\tabcaption{\label{Tab:002}Renormalization factors $N_{\rm R}$ for the real parts of the OPs used in CDCC calculations of $^6$Li+$^{12}$C.}
\footnotesize
\begin{tabular}{cc}
\hline
$E/A$(MeV) & $N_{\rm R}$\\
\hline
28.1 & 0.85 \\
35 and 53 & 0.9 \\
\hline
\end{tabular}
\end{center}

\vskip -2mm
\begin{center}
\includegraphics[width=\linewidth]{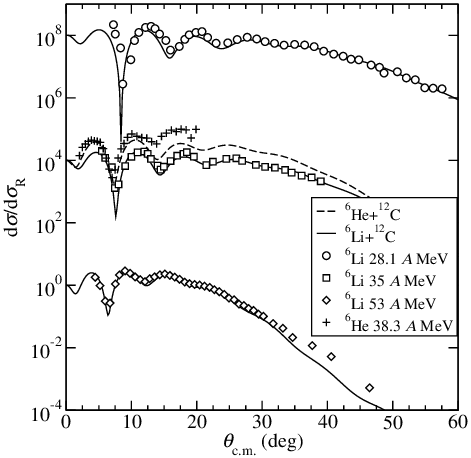}
\figcaption{\label{fig:003}Comparisons between CDCC calculations and experimental ESAD data of $^6$Li+$^{12}$C~\ref{b31,b32,b33} and $^6$He+$^{12}$C~\cite{b34} at different energies. The data set at 53\,$A$\,MeV is on the real scale, and the others are offset by factors of 10000.}
\end{center}

The improved dineutron cluster model of $^6$He given by Moro {\it et al.}~\cite{b1} is used for $^6$He+$^{12}$C. The BPs for the ground state, the non-resonant continuum and the $l$=2 resonance are all taken from Ref.~\cite{b1}. 
Considering the similarities between the structures of $^6$He and $^6$Li, the calculation settings similar to those of $^6$Li+$^{12}$C are applied to the calculation of $^6$He+$^{12}$C. Particularly, $N_{\rm R}$=0.9 is assumed, which is analogous to $^6$Li+$^{12}$C at 35\,$A$\,MeV.
Compared with the experimental ESAD, the calculation strongly underestimates the elastic differential cross section at angles larger than 10 degrees (Fig.\,\ref{fig:003}). This means that the calculation considering the breakup couplings overestimates the absorption of $^4$He+$^{12}$C heavily.
This phenomenon has been reported in Ref.~\cite{b17}, where the OP used for $^4$He+$^{12}$C is derived from the experimental ESAD at an adjacent energy 34.75\,$A$\,MeV~\cite{b22}. With the CDCC calculation using the COP, we have confirmed this phenomenon. 
Moreover, as an unstable halo nucleus, $^6$He was expected to cause a stronger absorption than $^6$Li when scattering on the carbon target at similar energies. However, $^6$He+$^{12}$C at 38.3\,$A$\,MeV has a larger elastic differential cross section (i.e. a weaker absorption) than $^6$Li+$^{12}$C at 35\,$A$\,MeV (Fig.\,\ref{fig:003}). Both the calculated results and experimental data have this behavior, which is also consistent with Ref.~\cite{b17}. 

In summary, the elastic scattering of $^4$He on $^{12}$C has been studied comprehensively with an OM. It is found that the ESAD is more sensitive to the real part than to the imaginary part of the optical potential. Meanwhile, the SPP considering the Pauli nonlocality includes the energy dependence naturally. Using a nuclear potential with the SPP as its real part and a fixed WS function as its imaginary part, the optical potential reproduces the ESADs of existing data within the energy range of 26-60\,$A$\,MeV.  It interprets an ultra weak energy dependence of the imaginary part in this energy range.  
The CDCC calculations by using the COP indicate the reliability of the COP in $^4$He-cluster involved breakup reactions. On the other hand, the puzzle of the weaker absorption of $^6$He+$^{12}$C than that of $^6$Li+$^{12}$C is confirmed.

We thank Thompson I. J., Moro A. M. and Pang D. Y. for their help in the usage of FRESCO. Thanks are also given to Chamon L. C. for his help in the consideration of the SPP.

\end{multicols}
\vspace{-1mm}
\centerline{\rule{80mm}{0.1pt}}
\vspace{2mm}

\begin{multicols}{2}

\end{multicols}

\clearpage

\end{document}